\documentclass{aa}
\usepackage{graphics,epsfig,txfonts,xspace}
\usepackage{natbib}
\bibpunct{(}{)}{;}{a}{}{,}

\newcommand{\ergs}[1]{$\times 10^{#1}$ erg s$^{-1}$}

\newcommand{\hcm}[1]{$\times 10^{#1}$ cm$^{-2}$}

\newcommand{\Hone}{\ion{H}{I}}

\newcommand{\ltsima}{$\buildrel < \over \sim$}
\newcommand{\lsim}{\lower.5ex\hbox{\ltsima}}
\newcommand{\gtsima}{$\buildrel > \over \sim$}
\newcommand{\gsim}{\lower.5ex\hbox{\gtsima}}
\newcommand{\msun}{M$_{\odot}$}
\newcommand{\xmm}{XMM-Newton\xspace}

\newcommand{\smcthree}{\mbox{SMC\,3}\xspace}

\usepackage{color}

\begin{document}
 
\title{The XMM-Newton survey of the Small Magellanic Cloud:\\ 
       A new X-ray view of the symbiotic binary \smcthree}
\author{R.~Sturm\inst{1} \and F.~Haberl\inst{1} \and J.~Greiner\inst{1}  \and W.~Pietsch\inst{1} \and 
        N.~La\,Palombara\inst{2} \and M. Ehle\inst{3} \and M. Gilfanov\inst{4} \and A. Udalski\inst{5} \and S. Mereghetti\inst{2}  \and M. Filipovi{\'c}\inst{6}}

\titlerunning{A new X-ray view of the symbiotic binary \smcthree}
\authorrunning{Sturm et al.}

\institute{Max-Planck-Institut f\"ur extraterrestrische Physik,
           Giessenbachstra{\ss}e, 85748 Garching, Germany
	   \and
	   INAF, Istituto di Astrofisica Spaziale e Fisica Cosmica Milano, via E. Bassini 15, 20133 Milano, Italy
	   \and
           XMM-Newton Science Operations Centre, ESAC, ESA, PO Box 50727, 28080 Madrid, Spain
	   \and
	   Max-Planck-Institut f\"ur Astrophysik, Karl-Schwarzschild-Str.1, 85741 Garching, Germany;\\
	   Space Research Institute, Russian Academy of Sciences, Profsoyuznaya 84/32, 117997 Moscow, Russia
           \and
           Warsaw University Observatory, Aleje Ujazdowskie 4, 00-478 Warsaw, Poland
	   \and
           University of Western Sydney, Locked Bag 1797, Penrith South DC, NSW1797, Australia
	   }

\date{Received 21 October 2010}

\abstract{The XMM-Newton survey of the Small Magellanic Cloud (SMC) was performed to study the population of X-ray sources in this neighbouring galaxy.
          During one of the observations, the symbiotic binary \smcthree was found at its highest X-ray luminosity observed until now.} 
         {In \smcthree wind accretion from a giant donor star onto a white dwarf is believed to cause steady hydrogen burning 
          on the white dwarf surface, making such systems candidates for supernova type Ia progenitors. 
	  It was suggested that the X-ray source is eclipsed every $\sim$4.5 years by the companion star and its stellar wind to 
	  explain the large X-ray variability seen in ROSAT data.
          We use the available  X-ray data to test this scenario.}
	 {We present the $\sim$20 year X-ray light curve of \smcthree and study the spectral evolution as seen with XMM-Newton/EPIC-pn to investigate 
	  possible scenarios which can reproduce the high X-ray variability.}
         {We did not find significant variations in the photo-electric absorption, as it would be expected during eclipse ingress and egress. 
	  Instead, the X-ray spectra from different intensity levels, when modelled by black-body emission, can be better explained by variations 
	  either in normalisation (by a factor of $\sim$50) or in temperature (kT between 24 eV and 34 eV). 
	  The light curve shows maxima and minima with slow transitions between them.}
         {To explain the gradual variations in the X-ray light curve and to avoid changes in absorption by neutral gas, a predominant part of the 
	  stellar wind must be ionised by the X-ray source. Compton scattering with variable electron column density (of the order of 5\hcm{24}) along the 
	  line of sight could then be responsible for the intensity changes. The X-ray variability of \smcthree could also be caused by temperature 
	  changes in the hydrogen burning envelope of the white dwarf, an effect which could even dominate if the stellar wind density is not 
	  sufficiently high.}

\keywords{stars: individual: SMC3 --
          (stars:) binaries: symbiotic --
          X-rays: binaries --
          (stars:) white dwarfs --
          galaxies: individual: Small Magellanic Cloud}
 
\maketitle
 
\section{Introduction}

The symbiotic star \smcthree \citep{1992MNRAS.258..639M} in the Small Magellanic Cloud (SMC) 
was discovered as super-soft source \citep[SSS, ][]{1994RvMA....7..129H} in X-rays during the ROSAT all sky survey \citep{1993LNP...416...71K}.
It is thought to be an interacting binary system, consisting of a cool M0 giant and a hot white dwarf (WD) in a wide orbit. 
In this model accretion from the stellar wind of the giant donor onto the WD leads to steady hydrogen burning on the WD surface 
which powers the high X-ray luminosity \citep{1997ARA&A..35...69K}.

A series of ROSAT observations covering $\sim$6 years (from Oct. 1990 to Nov. 1993 with the PSPC and from 
Apr. 1994 to Nov. 1996 with the HRI), revealed high X-ray variability (a factor of $\geq$80 in ROSAT PSPC count rate),
which was explained by an eclipse of the WD by the donor star \citep{2004A&A...416...57K}. 
This scenario needs obscuration of the X-ray emission region by the dense stellar wind close to the giant to account for the shape 
and the long duration of several months of the eclipse ingress and egress.

An optical outburst between December 1980 and November 1981 of up to 3 mag in the U band was reported by \citet{1992MNRAS.258..639M}. 
During this outburst no changes were detected in the I band and therefore, its origin was assigned to the hot stellar component.
It is not clear, if the non-detection of \smcthree with the Einstein satellite  was due to X-ray inactivity before the optical outburst 
or to insufficient sensitivity \citep{2004A&A...416...57K}.
The enrichment of nitrogen also suggests evidence for a thermonuclear event \citep{1994A&A...288..842V}.
Results from modelling multi-wavelength data of \smcthree with non-LTE models under the assumption of a constant X-ray source
 were presented in \citet{2007ApJ...661.1105O} and \citet{1996A&A...312..897J}.
\citet{2007ApJ...661.1105O} found that the variability cannot be caused by photo-electric absorption 
and suggested a ''real'' eclipse by the red giant.

\smcthree was in a very luminous state during observation of field number 13 of the 
\xmm \citep{2001A&A...365L...1J} large program SMC survey \citep{2008xng..conf...32H}.
This enables spectral analysis of the EPIC-pn \citep{2001A&A...365L..18S} data with unprecedented statistical quality.
Four spectra from different intensity states allow us now to study 
the spectral evolution of the hot component of \smcthree.
We present the light curve starting with the first ROSAT detection in 1990, 
to investigate the nature of the variability seen from this system.

\section{Observations and data reduction}

\begin{table*}
\caption{\xmm EPIC-pn observations of \smcthree}
\begin{center}
\begin{tabular}{lclcclrrrrrr}
\hline\hline\noalign{\smallskip}
\multicolumn{1}{c}{ObsID} &
\multicolumn{1}{c}{Satellite} &	
\multicolumn{1}{c}{Date} &	
\multicolumn{1}{c}{Time} &
\multicolumn{1}{c}{Filter} &
\multicolumn{1}{l}{Net Exp} &   
\multicolumn{1}{r}{Net cts.} &	
\multicolumn{1}{r}{Bg$^{(a)}$} &
\multicolumn{1}{r}{Net cts.} & 
\multicolumn{1}{r}{Bg$^{(a)}$} &
\multicolumn{1}{r}{R$_{\rm sc}^{(b)}$}&
\multicolumn{1}{r}{R$_{\rm bg}^{(b)}$}\\
\multicolumn{1}{l}{} &
\multicolumn{1}{l}{Revolution} &
\multicolumn{1}{l}{} &	
\multicolumn{1}{c}{(UT)} &
\multicolumn{1}{l}{} &	
\multicolumn{1}{c}{[s]} &   
\multicolumn{2}{c}{(0.2$-$10.0 keV)} &	
\multicolumn{2}{c}{(0.2$-$1.0 keV)} &
\multicolumn{1}{r}{[\arcsec]}&
\multicolumn{1}{r}{[\arcsec]}\\
\noalign{\smallskip}\hline\noalign{\smallskip}
0301170501 & 1149 &   2006 Mar 19    &	14:45-20:17 & medium   &  10446$^{(c)}$  &  8287	& 12\% &  8246  & 2\% & 19&30	\\
0404680301 & 1344 &   2007 Apr 11-12 &	20:00-02:15 & thin     &  13986          &   170	& 13\% &   137  & 5\% & 16&50	\\
0503000201 & 1444 &   2007 Oct 28    &	06:11-11:52 & medium   &  16607          &   451	& 9\%  &   407  & 4\% & 21&40	\\
0601211301 & 1798 &   2009 Oct 3     &	05:31-14:12 & thin     &  26535          & 39456	& 2\%  & 39338  & 1\% & 69&55	\\
\noalign{\smallskip}\hline\noalign{\smallskip}
\multicolumn{10}{l}{\hbox to 0pt{\parbox{180mm}{\footnotesize
\smallskip
$^{(a)}$ratio of background count rate to source count rate in the same energy band.\\
$^{(b)}$radius of the source and background extraction region.\\
$^{(c)}$no GTI screening was applied.\\

}}}
\end{tabular}
\end{center}
\label{tab:xray-obs}
\end{table*}

\smcthree was serendipitously observed four times with \xmm at off-axis angles between 8\arcmin\ and 14\arcmin.
Table~\ref{tab:xray-obs} lists some details of the observations with the EPIC instruments operated in full-frame mode.
In addition to the observation in Oct. 2009 from the SMC large survey program, we analysed three observations from 
2006 and 2007 available in the archive.
The first observation in March 2006 revealed \smcthree in a high intensity state, but suffered from very high background. 
These data were used in the study of \citet{2007ApJ...661.1105O}.
The two observations in April and October 2007 showed the source at low intensity.
The detection of the source in the later observation was noted by \citet{2008ATel.1379....1Z}.

\begin{figure}
 \resizebox{1.\hsize}{!}{
 \includegraphics[height=6cm,angle=0,clip=]{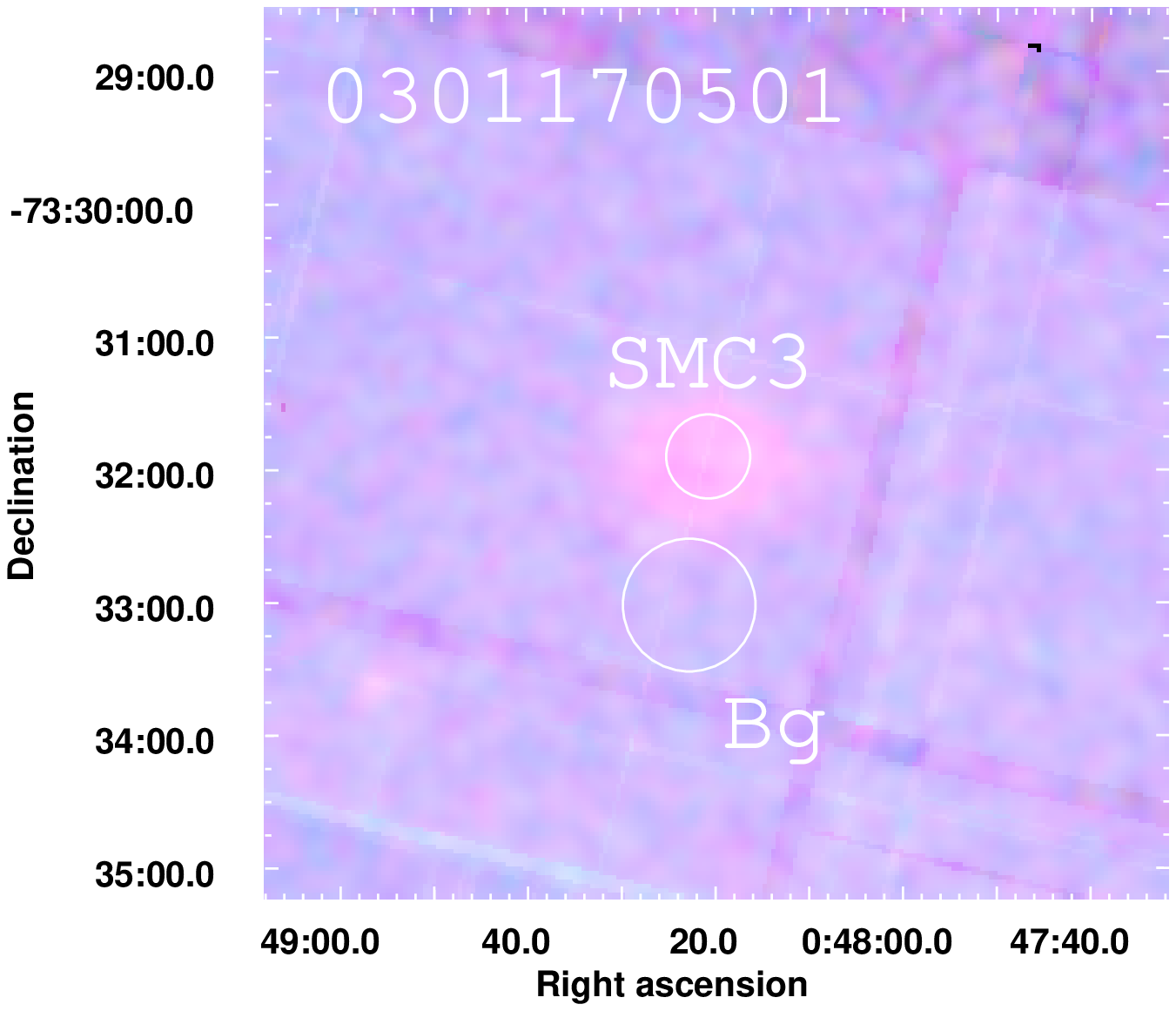}\quad
 \includegraphics[height=6cm,angle=0,clip=]{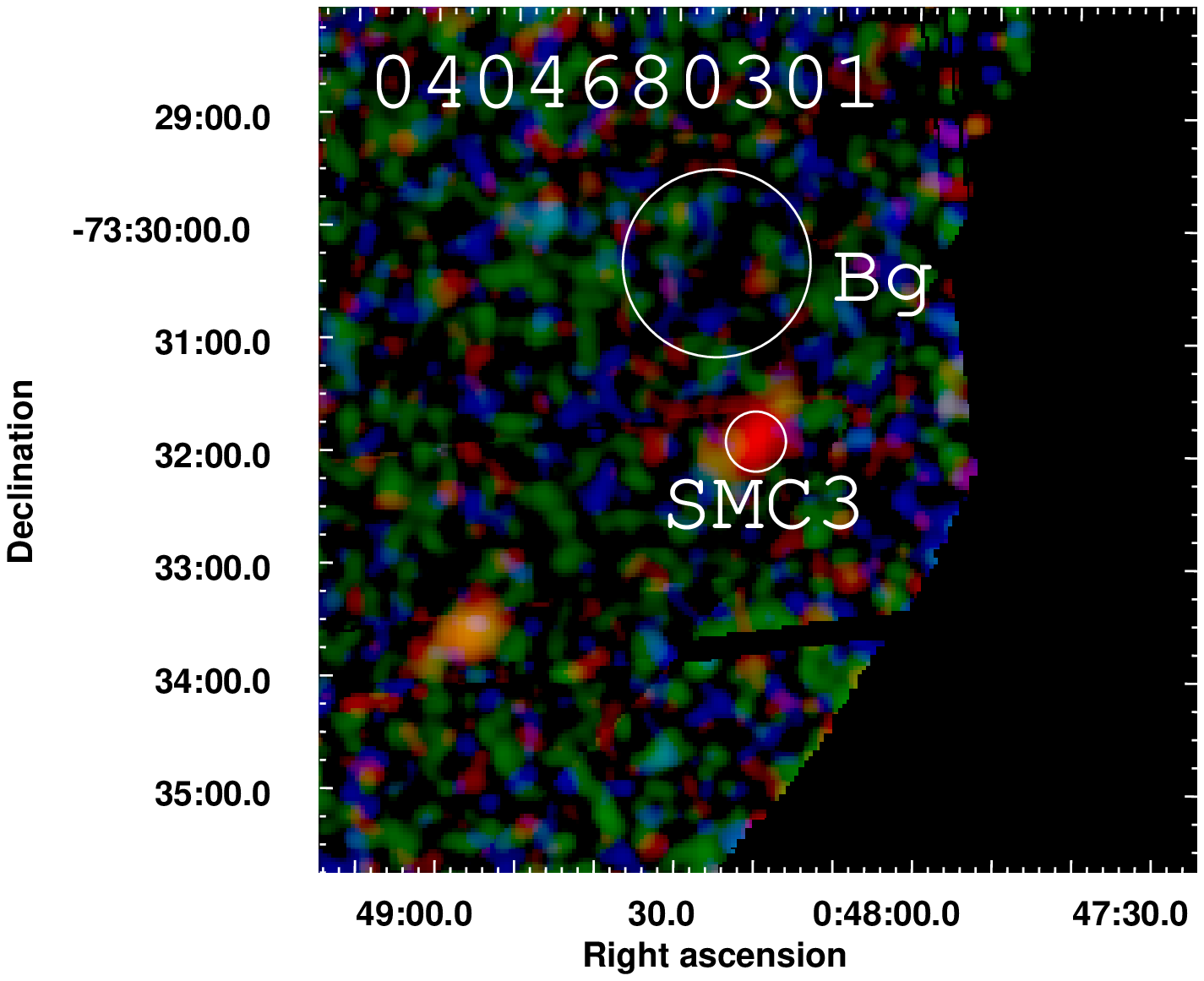}
}
\newline 
 \resizebox{1.\hsize}{!}{
 \includegraphics[height=6cm,angle=0,clip=]{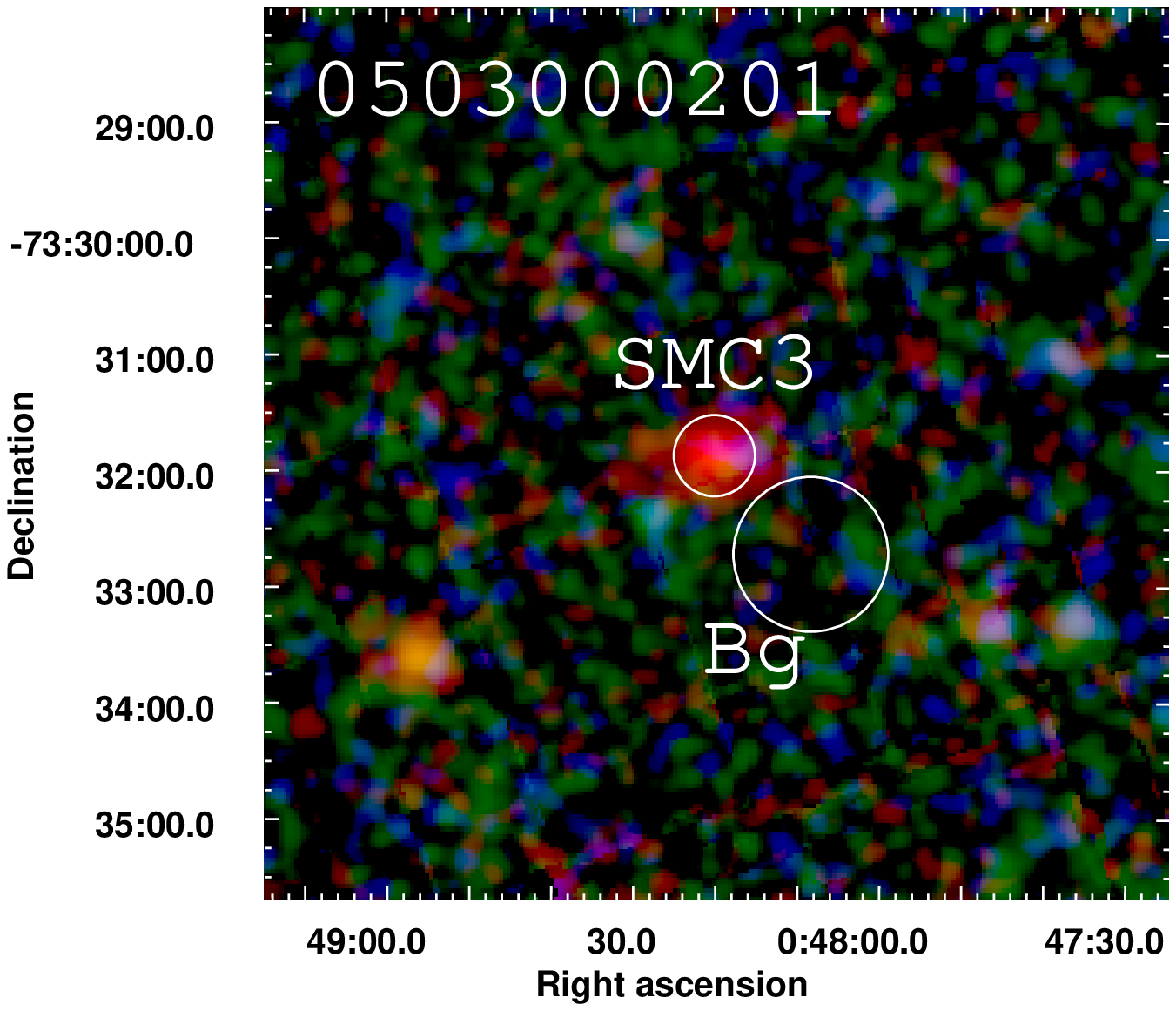}\quad
 \includegraphics[height=6cm,angle=0,clip=]{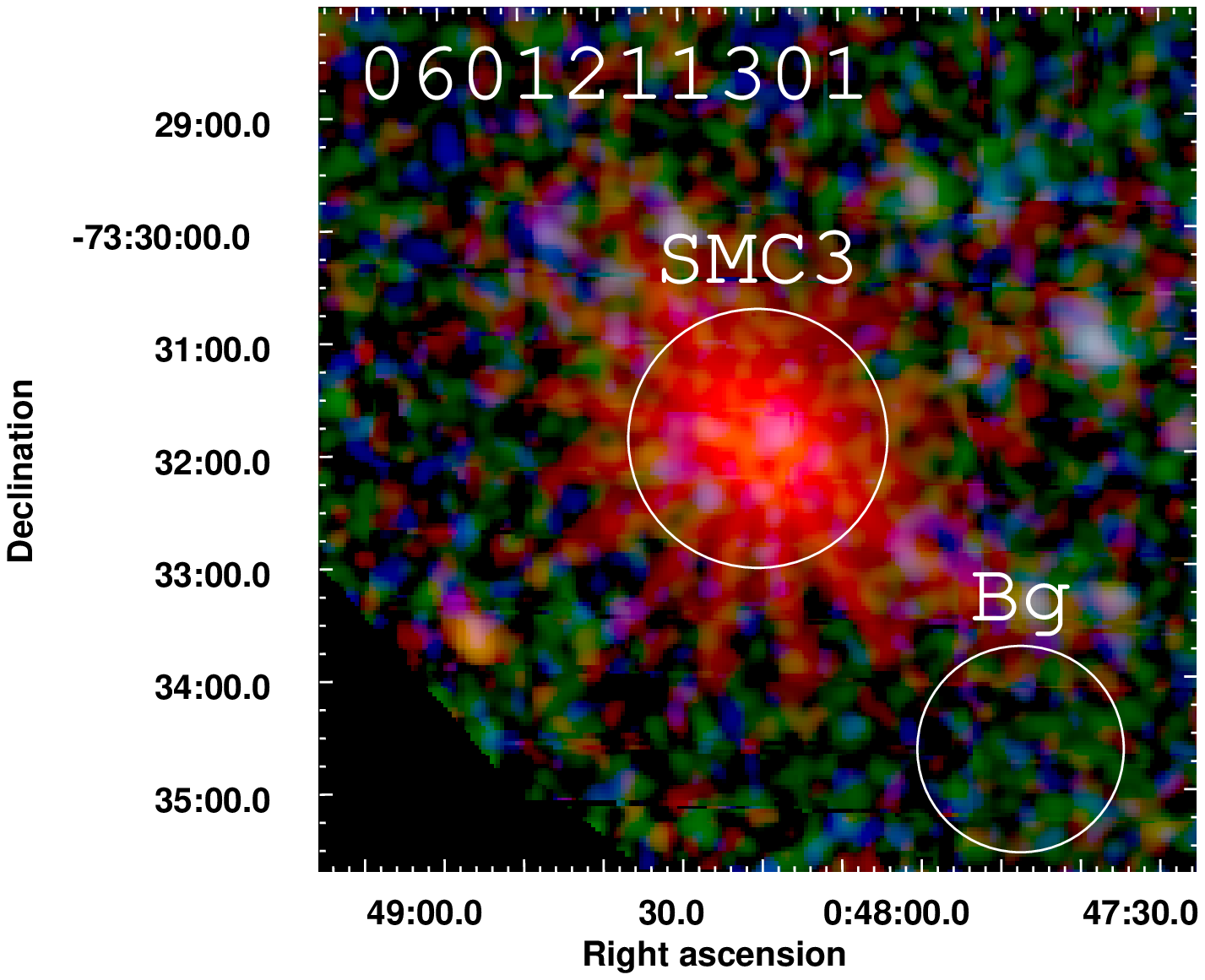}
}
  \caption{Combined EPIC colour images of \smcthree from the four XMM-Newton observations. Red, green, and blue colours denote X-ray 
           intensities in the 0.2$-$1.0, 1.0$-$2.0 and 2.0$-$4.5 keV bands. Circles indicate the extraction regions. Note the high 
	   background during the observation shown in the upper left which strongly contributes to higher energies.}
   \label{fig:ima}
   
\end{figure}

To process the data, we used XMM-Newton SAS 10.0.0\footnote{Science Analysis Software (SAS), http://xmm.esac.esa.int/sas/} 
with calibration files available until 17 June 2010, including the latest refinement of the EPIC-pn energy redistribution. 
As a standard, we selected good time intervals (GTIs) with an EPIC-pn background rate below 8 cts ks$^{-1}$ arcmin$^{-2}$ 
(single- and double-pixel events, 7$-$15 keV). However, for the observation in 2006 the background was between 500 and 
2500 cts ks$^{-1}$ arcmin$^{-2}$ and in order to retain any data, no background screening was applied. 
Since soft proton flares usually show a rather hard spectrum, the contribution to the super-soft spectrum of 
\smcthree was still acceptable (cf. Table~\ref{tab:xray-obs}).
The SAS task {\tt eregionanalyse} was used to determine circular extraction regions by optimising the signal to noise ratio, 
as shown in Fig.~\ref{fig:ima} and listed in Table~\ref{tab:xray-obs}. 
We ensured that the source extraction region had a distance of $>$10\arcsec\ to other detected sources.
For the background extraction region, we chose a circle on a point source free area on the same CCD as the source.
Since the MOS-spectra have lower statistical quality by a factor of 10 for such soft spectra and to avoid cross calibration 
effects between the EPIC instruments\footnote{EPIC Calibration Status Document,\\ http://xmm2.esac.esa.int/external/xmm\_sw\_cal/calib/index.shtml}, 
we are concentrating on the EPIC-pn spectra in this study.
For the extraction of EPIC-pn spectra, we selected single-pixel events with {\tt FLAG = 0}. We binned the spectra to a minimum 
signal-to-noise ratio of 5 for each bin using the task {\tt specgroup}.

\section{Spectral analysis of the EPIC-pn data}

\begin{figure}
  \resizebox{\hsize}{!}{\includegraphics[angle=-90]{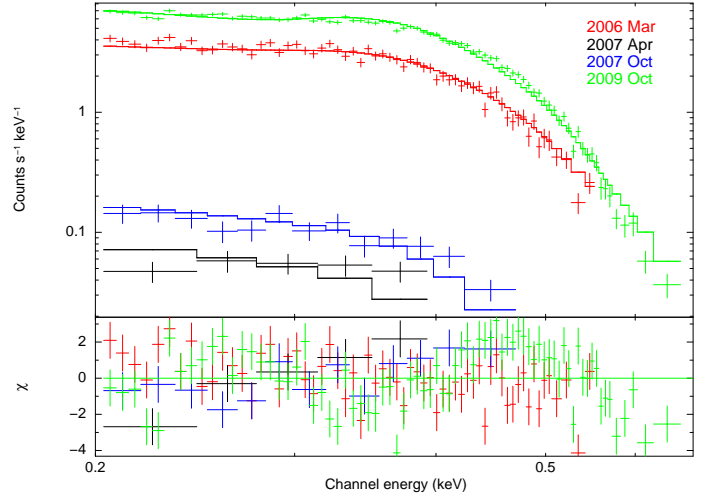}}
  \caption{EPIC-pn spectra of \smcthree together with the best-fit black-body model 3 with variable temperature. 
           No significant emission is seen above 0.7 keV.}
  \label{fig:spec}
\end{figure}

We used {\tt xspec} \citep{1996ASPC..101...17A} version 12.5.0x for spectral fitting.
For all models, the Galactic photo-electric absorption was fixed 
at a column density  of N$_{\rm H{\rm , gal}}$ = 6\hcm{20} and elemental abundances according to \citet{2000ApJ...542..914W},
whereas the SMC column density was a free parameter with abundances at 0.2 for elements heavier than Helium.
At first, we investigated the recent EPIC-pn spectrum of 2009, which has unprecedented statistics compared to previous X-ray observations. 
For black-body emission, we obtain a best-fit with $\chi^2/{\rm dof} = 213/74$ with the best-fit parameters:
N$_{\rm H, SMC} = 7.84^{+0.44}_{-0.24}$ \hcm{20}, $kT = 33.9\pm0.5$ eV, and
a bolometric luminosity of $L_{bol}=6.25^{+1.08}_{-0.86}$ \ergs{38}.
This luminosity is super-Eddington (and leading to a WD radius $\sim 5-10$ times larger than expected for a 1 \msun\ WD), 
which is often caused by the black-body approximation 
\citep[see e.g.][ and references therein]{1991A&A...246L..17G,1997ARA&A..35...69K}.
The residuals around 500 eV (cf. Fig.~\ref{fig:spec}) suggest contribution of nitrogen line emission, as observed in post-nova X-ray spectra of SSS \citep[e.g. ][]{2009AJ....137.4627R}.
The quality of the fit was improved to $\chi^2/{\rm dof} = 119/72$ by including two Gaussian lines with fixed energy at 431 eV (\ion{N}{vi}) 
and 500 eV (\ion{N}{vii}) and line widths fixed to 0. 
The equivalent widths of 19 eV and 32 eV for the \ion{N}{vi} and \ion{N}{vii} lines, 
respectively, are physically plausible, but the residuals could at least partially be also caused by calibration uncertainties.
Alternatively, allowing the oxygen abundance in the SMC absorption component as free parameter also improved the fit ($\chi^2/{\rm dof} = 123/73$), 
resulting in an oxygen abundance of $9.9^{+4.5}_{-2.7}$ times solar with an N$_{\rm H, SMC} = 4.39^{+0.66}_{-0.79}$ \hcm{20}.
A hard spectral component, which could possibly be caused by the wind-nebula, is not seen in the spectrum. 
Adding an apec plasma emission component to the black-body emission, with fixed temperature of $kT=500$ eV and SMC-abundances 
yields an upper limit for the emission measure of $EM=8.6 \times 10^{57}$ cm$^{-3}$.

The derived values for the absorption N$_{\rm H, SMC}$ are well below the total SMC absorption in the direction of \smcthree 
\citep[$\sim$5\hcm{21}; see ][]{1999MNRAS.302..417S}. This suggests that the symbiotic system is located on the near side of the large 
amount of \Hone\ present in the SMC Bar.

We also tested non-local thermal equilibrium models provided by 
Thomas Rauch\footnote{http://astro.uni-tuebingen.de/$\sim$rauch/} \citep{2010AN....331..146R}. 
We found the best-fit for a pure He atmosphere with WD surface gravity 
log g = 5 ($\chi^2/{\rm dof} = 220/74$).
For a pure hydrogen atmosphere with log g = 9 we found a best-fit at $\chi^2/{\rm dof} = 261/74$.
If the spectrum in fact contains emission lines, this might be the reason of the inferior fit of the non-LTE models which produce spectra 
which are dominated by absorption lines from the white dwarf atmosphere. 
Since the black-body spectrum resulted in a better fit, we decided to use this model in our further investigations.

To study the spectral evolution of \smcthree, 
we fitted the EPIC-pn spectra of all four epochs simultaneously with a set of models based on absorbed black-body emission. In each model
only one individual parameter for each spectrum and two common parameters for all spectra were allowed to vary (see below).
Since the statistical quality of the high-flux spectrum is far better than for the other spectra, it also dominates 
the resulting $\chi^2$. This leads to relatively bad fits for the simple black-body models. Adding emission lines would improve 
the fits (see above). However, given the limited spectral resolution of the EPIC-pn instrument it is not clear if the lines have any 
physical meaning. Because we were mainly interested in the evolution of spectral parameters, we decided to use the 
simplest model.

Our first model (model 1 in Table~\ref{tab:bbfit}) assumes temperature and luminosity not varying with time, 
while the absorbing column density can change with time.
This corresponds to the eclipse model with varying absorption by the dense donor wind, as suggested by \citet{2004A&A...416...57K}.
This model gives an insufficient fit to the data (see Table~\ref{tab:bbfit}).
Although, the fit is statistically dominated by the two high-flux spectra, the spectral shape of the low-flux 
spectra cannot be reproduced by a high column density which predicts much less flux at lowest energies. 

For model 2, we fixed the spectral shape (same temperature and absorption) and allowed the luminosity to change (i.e. fitting 
individual normalisations which corresponds to a variable size of the emission area). This fits the data much better (Table~\ref{tab:bbfit}).

An even better fit was achieved by our third model with varying source temperature. The corresponding black-body luminosities 
were related to the temperature ($L_{\rm bol} \propto T^4$).
The results for this model are again described in Table~\ref{tab:bbfit} and the individual spectra with the model fit are plotted 
in Fig.~\ref{fig:spec}.

\begin{table*}
\caption{Results from the simultaneous black-body fit to the EPIC-pn spectra.}
\begin{center}
\begin{tabular}{lll}
\hline\hline\noalign{\smallskip}
\multicolumn{1}{c}{Model 1} &
\multicolumn{1}{c}{Model 2} &
\multicolumn{1}{c}{Model 3} \\
\multicolumn{1}{c}{$T$ and $L_{\rm bol}$ constant with time} &
\multicolumn{1}{c}{N$_{\rm H}$ and $T$  constant with time} &
\multicolumn{1}{c}{N$_{\rm H}$  constant with time} \\	
\multicolumn{1}{c}{N$_{\rm H}$  variable} &
\multicolumn{1}{c}{$L_{\rm bol}$  variable} &	
\multicolumn{1}{c}{$T$ variable, $L_{{\rm bol,} i}= L_{\rm bol, 1} (T_i/T_1)^4$} \\	

\noalign{\smallskip}\hline\noalign{\smallskip}

$kT= (32.5 \pm 0.5)$ eV                              & N$_{\rm H} = (0.77 \pm 0.04) \times 10^{21}$ cm$^{-2}  $   & N$_{\rm H} = (0.77 \pm 0.02) \times 10^{21}$ cm$^{-2}  $\\
\vspace{1mm}
$L_{\rm bol}= (8.8 \pm 1.6) \times 10^{38}$ erg s$^{-1}$    & $kT= (33.7 \pm 0.5)$ eV                                 & $L_{\rm bol, 1} = (5.5 \pm 0.3) \times  10^{38}$ erg s$^{-1}$ \\
N$_{\rm H, 1} = (1.08\pm0.05) \times 10^{21}$ cm$^{-2}$  & $L_{\rm bol, 1} = (4.43 \pm 0.78) \times 10^{38}$ erg s$^{-1}$ & $kT_1=(32.7 \pm 0.2)$ eV  \\
N$_{\rm H, 2} = (6.75\pm0.48) \times 10^{21}$ cm$^{-2}$  & $L_{\rm bol, 2} = (0.12 \pm 0.02) \times 10^{38}$ erg s$^{-1}$ & $kT_2=(24.3  \pm 0.3)$ eV \\
N$_{\rm H, 3} =  (6.91\pm0.23) \times 10^{21}$ cm$^{-2}$ & $L_{\rm bol, 3} = (0.18 \pm 0.03) \times 10^{38}$ erg s$^{-1}$ & $kT_3=(25.3 \pm  0.2) $ eV\\\vspace{1mm}
N$_{\rm H, 4} = (0.81\pm0.05) \times 10^{21}$ cm$^{-2}$  & $L_{\rm bol, 4} = (6.47 \pm 1.15) \times 10^{38}$ erg s$^{-1}$ & $kT_4=(33.8 \pm 0.4) $ eV \\
$\chi^2/{\rm dof}= 940/154 = 6.10$                       & $\chi^2/{\rm dof}= 383/154 = 2.49$                             & $\chi^2/{\rm dof}= 365/154 = 2.37$  \\
\noalign{\smallskip}\hline\noalign{\smallskip}
\end{tabular}
\end{center}
\label{tab:bbfit}
\end{table*}

\section{The X-ray light curve of \smcthree}

To analyse the temporal behaviour of the system, we reconciled the X-ray light curve of \smcthree starting from the first detection by ROSAT in 1990.
To convert the ROSAT count rates, provided by \citet{2004A&A...416...57K}, 
into fluxes, we simulated a ROSAT PSPC spectrum based on the spectral model derived from the simultaneous fit with variable normalisation 
and the PSPC detector response. We obtained a conversion factor of $1.54 \times 10^{-11}$ erg cm$^{-2}$ cts$^{-1}$. 
All fluxes are computed for the 0.2$-$1.0 keV band.
Because of the dominant statistics of the XMM-Newton high-flux spectrum in the model fit, this factor rather corresponds to the high-flux state. 
Using the model with variable temperature, the conversion factor for the low flux can be lower by a factor of $\lesssim$2.
Analogous, for the ROSAT HRI, this simulation yields a conversion factor of $4.63 \times 10^{-11}$ erg cm$^{-2}$ cts$^{-1}$.

We searched the Swift archive for observations covering \smcthree. The source was detected in observation 00037787001 
with $\sim$3~ks exposure on 18 August 2008. The source and background spectrum was extracted with {\tt xselect} and the 
effective area file was created using {\tt xrtmkarf}.
The resulting spectrum contains 135 net counts, which is insufficient to distinguish between the above described models. 
Thus we fitted only the normalisation and assumed the spectral shape according to  the simultaneous fit to the EPIC-pn spectra with variable normalisation.
This fit yields a flux of $2.61^{+0.26}_{-0.15} \times 10^{-12}$ erg cm$^{-2}$  s$^{-1}$.

To derive XMM-Newton fluxes, we integrated the best-fit model with variable temperature as described above.
The source flux during a Chandra observation in February 2003 was deduced from the parameters of the best-fit black-body model reported by \citet{2007ApJ...661.1105O}.

Figure~\ref{fig:lc} shows the X-ray light curve of \smcthree since the first detection by ROSAT in 1990.
By modelling the light curve with several eclipses, we realised,
that (i) the transition from high to low intensity occurs over a long time period and 
that (ii) the time scales of the duration of the high and low intensity intervals are of the same order.
Thus, instead of eclipses the light curve may also be interpreted by several periodic outbursts.
For demonstration, the dashed line in Fig.~\ref{fig:lc} shows a fit with a sine function. To account for 
uncertainties in the flux conversion and cross calibration between the different instruments we included 
a 20\% systematic error on the flux. As best-fit ephemeris for the X-ray minimum we then obtain (90\% confidence errors):
$${\rm MJD_{min,x} = (49382\pm 10) + N \times (1634 \pm 7)~days.}$$
The relatively high flux measured in the last \xmm observation might suggest possible changes in the amplitude of the modulation.

\section{MACHO and OGLE data}

\begin{figure}
  \resizebox{\hsize}{!}{\includegraphics[angle=-90]{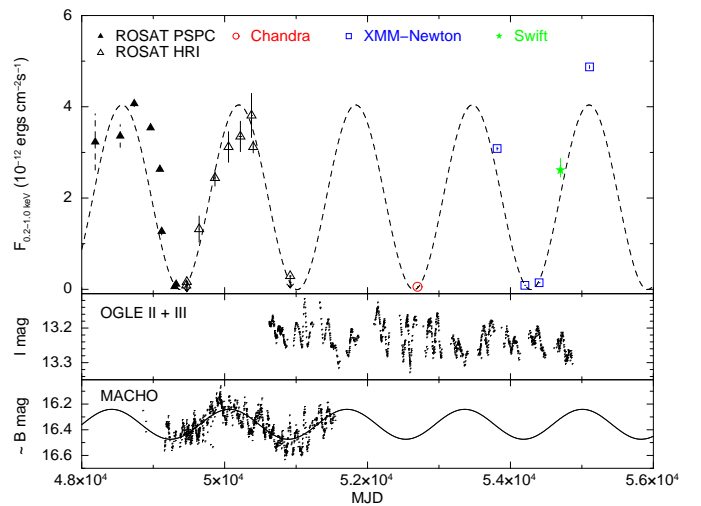}}
  \caption{The 0.2$-$1.0 keV X-ray light curve of \smcthree (upper panel), the light curves in the I-band (OGLE, middle panel) and the 
           approximate B-band (MACHO, lower panel). 
           The dashed line shows the best-fit sine function to the X-ray light curve and the solid line the best-fit sine function to 
	   the MACHO light curve (see text).}
  \label{fig:lc}
\end{figure}

The OGLE-II \citep{1997AcA....47..319U} and OGLE-III \citep{2008AcA....58...69U} I-band as well as the 
MACHO B-band light curves are shown in the lower two panels of Fig.~\ref{fig:lc}.
The calibrated MACHO light curve was shifted in magnitude to match its average B magnitude with that measured by \citet{2002AJ....123..855Z}.
In the I-band of OGLE-II and the B-band of MACHO, \citet{2004A&A...416...57K} found correlating quasi-periodic oscillations with periods around 
110 days which might be related to pulsations of the red giant star.
These short variations are also present in the OGLE-III data. With the OGLE-III data, which cover a much longer time interval, 
we can now rule out a significant variation with the 1630 day cycle as suggested by the X-ray light curve.
As noted by \citet{2004A&A...416...57K}, the MACHO light curve shows a quasi sinusoidal modulation with a period of $\sim$4 years, 
in addition to shorter variations.
A fit of a sine function to the MACHO B-band data (solid line in Fig.~\ref{fig:lc}), results in an ephemeris for the optical minimum of
$${\rm MJD_{min,B} = (49242\pm 9)  + N \times (1647 \pm 24)~days.}$$
To account for the short-term variations we added a systematic error of 0.05 mag to the B-band magnitudes.
Formally, the fits to X-ray and MACHO light curves indicate a phase shift of $(140\pm14)$ days (optical preceding the X-rays) while the periods agree within the errors.
However, it should be noted that the MACHO light curve covers only $\sim$1.5 cycles and is superimposed by the short term variations which may 
influence the results.

\section{Discussion}

\smcthree was observed with \xmm at four epochs, covering the super-soft X-ray source twice at high and twice at low intensity.
A strong variation in the X-ray flux by a factor of more than $\sim$50 between minimum and maximum intensity is found in the 0.2$-$1.0 keV band.
We showed, that the light curve can qualitatively be described by a sine function with a period of 1633 days.
This simple model can only be a crude approximation of the light curve, but the $\sim$20 year coverage indicates a high 
regularity of the period with similar duration of high and low intensity intervals. The regularity of the X-ray light curve, with meanwhile four observed minima, 
strongly supports the interpretation of the 4.5 year period as the orbital period of the binary system.
Assuming masses of 15 \msun\ and 1 \msun\ for the M-giant and the white dwarf, respectively, the orbital period implies a semi-major axis of the 
binary system of 6.83\,AU. 

We analysed the spectral evolution and found, that the variability of the X-ray flux cannot be explained by photo-electric absorption 
by neutral gas with varying column density. To avoid the strongly energy-dependent attenuation of soft X-rays, \citet{2004A&A...416...57K} 
discussed absorption due to highly ionised gas. In this picture the strong X-ray source ionises the stellar wind around it. 
Compton scattering on free electrons would then reduce the X-ray flux along the line 
of sight most efficient when looking through the dense innermost regions near the M-star. 
This mimics variable intensity with little energy dependence (no significant change of spectral shape).
Using a Compton scattering model ({\tt cabs} in {\tt xspec}), instead of variable normalisation, would require a column density of 
$>$4.8 \hcm{24} (completely ionised absorber) to reduce the X-ray intensity from maximum to minimum. 
In this picture scattering of X-rays into the line of sight is neglected or at least assumed not to change significantly between 
the two states. Using the estimated mass, size and density 
of the ionised wind region as given by \citet{2007ApJ...661.1105O} yields a column density to its centre of 5.8 \hcm{23}. This is a factor 
of $\sim$8 lower than our estimate from Compton scattering and may be explained by the simplified assumption of a constant wind density while the line of sight during 
the low intensity observations should pass through the denser wind regions near the giant star. In this picture most of the stellar wind 
must be ionised, consistent with the fact that we do not see a variable contribution of photo-electric absorption by neutral gas.
In this model, the low intensity can still be explained by an eclipse of the X-ray source by the giant star {\it and} its stellar wind. 
An eclipse by the star only would be short ($\sim$3\% of the orbital period) with sharp ingress and egress while the dense inner 
wind regions cause a long gradual eclipse ingress (and egress) by increasing (decreasing) Compton scattering along the line of sight. 
The shape of the light curve should then depend on the geometry of the binary 
system and the distribution of free electrons in the stellar wind.

Our spectral analysis shows, that the X-ray variability can alternatively be dominated by temperature changes, varying 
between 24 and 34 eV. Assuming a constant size of the emitting area, this corresponds to a variation in $L_{\rm bol}$ (for spherically 
symmetric emission) by a factor of 4.3. The larger variation in observed instrumental count rates would then be caused by shifting 
the spectra with lower temperature out of the sensitive energy band of EPIC.
A possible scenario might be an elliptical orbit of the white dwarf around the M-giant (or equatorial mass ejection with inclined WD orbit), 
causing accretion at different rates.
Variable accretion, even at low level, can lead to large temperature changes in the burning layer \citep{1980A&A....82..349P}.
Similar scenarios were used to explain X-ray variability in other SSS \citep[e.g. AG Dra, ][]{1996LNP...472..267G}. However, 
we note, that in those cases usually an anti-correlation of X-ray and optical luminosity is observed, 
whereas in the case of \smcthree these two are clearly correlated.

Assuming the same temperature and absorption (and no change in Compton scattering) for the low and high intensity spectra, 
the inferred radii would be different by a factor of $\sim$8 to account for the factor of 60 difference in $L_{\rm bol}$. 
In general for stable shell burning on the WD surface, an increase of the hydrogen burning envelope (e.g. due to a higher accretion rate) 
leads to an increase of both, temperature and radius \citep{1982ApJ...257..752F}. 
Increasing temperature and declining Compton scattering both lead to an increasing X-ray luminosity. 
It depends on the orientation of the orbit with respect to the observer, how much the two 
effects act in phase. Additional temperature variations 
may therefore reduce or increase the amount of Compton scattering required to explain the X-ray luminosity variations.

Superimposed on the general long term variation in the X-ray light curve, 
we probably see effects imposed by the donor star. The $\sim$110 days brightness variations seen in the I-band suggests 
changes in the stellar wind which can lead to variations in the mass accretion rate onto the white dwarf.
Since the 1630 day X-ray period is not visible in the I-band, it is unlikely that this period is caused by the cool stellar component, 
which seems to remain rather unaffected by the process producing this variation.
In the B-band, both modulations are seen, the 1630 day period derived from the X-rays and the $\sim$110 day variations 
which correlate with the I mag \citep[as already pointed out by ][]{2004A&A...416...57K}.
Therefore, the cool companion star and the region where the X-ray emission is produced most likely both contribute to the B-band.
If viewing effects produce the variation in B in a similar way as in the X-rays (by changing extinction) or if heating of the 
cool star by the X-ray source is causing this variation remains unclear: While \citet{2004A&A...416...57K} finds X-ray heating insufficient to account for the observed 
B-band modulation, \citet{2007ApJ...661.1105O} discuss irradiation effects influencing the mass outflow as very important.

\section{Conclusions}

The new X-ray data of \smcthree show that two scenarios can qualitatively explain the spectral evolution and the shape of the 
light curve. The evolution of the X-ray spectra is incompatible with changing photo-electric absorption by neutral gas, but is 
consistent with energy-independent intensity and/or with temperature variations. As suggested before by \citet{2004A&A...416...57K}, 
Compton scattering in a predominantly ionised stellar wind could lead to the observed intensity 
variations if the stellar wind density (mass loss rate) is high enough. Additional temperature changes in the burning layer of the 
WD which are caused by variable accretion, can reduce the required wind densities.

\begin{acknowledgements}
This publication is partly based on observations with XMM-Newton, an ESA Science Mission with instruments and contributions 
directly funded by ESA Member states and the USA (NASA).
The XMM-Newton project is supported by the Bundesministerium f\"ur Wirtschaft und 
Technologie/Deutsches Zentrum f\"ur Luft- und Raumfahrt (BMWI/DLR, FKZ 50 OX 0001)
and the Max-Planck Society. 
R.S. acknowledges support from the BMWI/DLR grant FKZ 50 OR 0907.
This paper utilises public domain data obtained by the MACHO Project, jointly
funded by the US Department of Energy through the University of California,
Lawrence Livermore National Laboratory under contract No. W-7405-Eng-48, by
the National Science Foundation through the Center for Particle Astrophysics
of the University of California under cooperative agreement AST-8809616, and
by the Mount Stromlo and Siding Spring Observatory, part of the Australian
National University.
\end{acknowledgements}

\bibliographystyle{aa}
\bibliography{../general}

\end{document}